\documentclass[prl,twocolumn,amssymb,amsmath]{revtex4-1}
\usepackage{graphicx}
\usepackage{epsfig,color}
\usepackage{amsmath,amssymb}
\begin{document}

\title{A cold-atom quantum simulator for SU(2) Yang-Mills lattice gauge theory}

\date{\today}

\author{Erez Zohar$^1$, J. Ignacio Cirac$^2$, Benni Reznik$^1$}

\begin{abstract}
\begin{center}
$^1$School of Physics and Astronomy, Raymond and Beverly Sackler
Faculty of Exact Sciences, Tel Aviv University, Tel-Aviv 69978, Israel.

$^2$Max-Planck-Institut f\"ur Quantenoptik, Hans-Kopfermann-Stra\ss e 1, 85748 Garching, Germany.
\end{center}
Non-abelian gauge theories play an important role in the standard model of particle physics, and
unfold a partially unexplored world of exciting physical phenomena.
In this letter, we suggest a realization of a non-abelian lattice gauge theory - SU(2) Yang-Mills
in 1+1 dimensions,  using
ultracold atoms.
Remarkably, and in contrast to previous proposals, in our model
gauge invariance is a direct consequence
of angular momentum conservation  and thus is
fundamental and robust.  Our  proposal may serve as well as a starting point for higher dimensional realizations.
\end{abstract}

\maketitle

The importance of gauge theories, being in the heart of the standard model of modern particle physics, cannot be over estimated. Gauge fields give rise to the long range causal interactions between matter particles. However, while Quantum Electrodynamics (QED) is an abelian gauge theory, the
strong interactions are described by Quantum Chromodynamics (QCD), a non-abelian $SU(3)$ Yang-Mills gauge theory.
This gives rise to many significant differences. For example, 
non-abelian theories involve the effect of \emph{quark confinement} \cite{Wilson,Polyakov}, which is responsible for the forces that "bind quarks together", forbidding the existence of free quarks.
This gives rise to the familiar structure of hadrons.
 This effect, as well as other non-perturbative phenomena of 3+1-d dimensional non-abelian models, required the development of new methods and techniques, such as lattice gauge theory \cite{Wilson,KogutSusskind,KogutLattice,Kogut1983}. It has been helpful to study fundamental QCD effects using simpler models that manifest the same essential ingredients.
 For example, confinement of quarks could be examined within the confining phase of (abelian) compact-QED \cite{Wilson,KogutSusskind,BanksMyersonKogut,DrellQuinnSvetitskyWeinstein}, which already in 2+1 dimensions gives rise to flux loops and plaquette interactions.
Another approach is to use the 1+1-d version of $SU(N_c)$ Yang-Mills theories (such as $QCD_2$, the 1+1-d version of QCD); 1+1 dimensional models have been used in numerous nonperturbative methods
to study the hadronic spectrum of such gauge theories \cite{tHooft1974,Callan1976,Witten1984,Frishman1993,Hornbostel1990,Gross1996,Armoni2001}.


The field of Quantum simulations \cite{Lewenstein2012,Bloch2012,Blatt2012} has been extensively developed theoretically and experimentally with the aim of advancing new quantum computational approaches to many body systems,  in particular in the context of condensed matter physics.  
Recently,  it has been realized that quantum simulations could  be used also for exploring High Energy Physics (HEP) models and effects \cite{CiracZoller}. 
Simulations of HEP models are in general more demanding, as compared with condensed matter ones for several reasons.
Matter and gauge fields are described by fermionic and bosonic fields, and thus the use of several atomic species is needed. Furthermore, the continuum limit of the models must include relativistic symmetries, which can be simulated with non-relativistic atoms by using lattices, as in lattice gauge theories.
There have been several suggestions for the simulation of relativisitic field theories involving bosons \cite{Retzker2005,Horstmann2010,Jordan2012}, fermions, free or interacting with external (classical) gauge fields \cite{Bermudez2010,Boada2011,Mazza2012}, or fermions interacting with (bosonic quantum) gauge fields in 1+1-d \cite{Cirac2010}.

The simulation of \emph{dynamic} gauge theories is even more challenging. First, gauge invariance must be respected.
 Moreover, it requires a special form of the many-body interactions, which are usually not available in most systems. In particular, lattice models involve plaquette interactions among the links of the lattice. Second, Gauss law (involving gauge bosons and fermions) has to be implemented as a constraint. The special interactions could be obtained as a low-energy effective gauge invariant theory of the original system.

 Very recently, realizations of abelian dynamic gauge theories, employing trapped atoms in optical lattices, have been proposed, simulating 2+1-d Kogut-Susskind (KS) abelian compact QED, that manifests confinement of charges, using either BECs \cite{Zohar2011} or single atoms (a truncated version) \cite{Zohar2012}.
These can also be extended to include dynamic matter, as proposed for the 1+1 dimensional Schwinger model \cite{Banerjee2012}, which can be compared to exact available solutions, and for a 2+1 dimensional truncated KS model \cite{Dynamic}. Simulations of other abelian gauge theories \cite{Kapit2011,Szirmai2011,Celi2012} have been proposed as well.

In this letter we present a simulation scheme for an $SU(2)$ Yang-Mills theory,  where both the fermions (matter) and gauge bosons are dynamical. The main idea is to introduce additional fermionic and bosonic fields (ancillas) to obtain the desired non-abelian aspect. Remarkably, and in contrast to previous proposals, in our model gauge invariance is fundamental and a direct consequence of angular momentum conservation, making this important property fundamental and robust.


\emph{Lattice Yang-Mills Theory}.
The system we wish to simulate includes a non-abelian gauge field and dynamic fermionic matter. In ordinary lattice theory the gauge field degrees of freedom are defined on the lattice's links, whereas the matter fields
are located on the vertices \cite{KogutSusskind,KogutLattice,Kogut1983}.
The gauge field is represented by unitary matrices $U_n^r$, whose
  elements $\left(U^r_n\right)_{kl}$ are constructed out of operators in the local (gauge field) Hilbert space. The index $n$ labels the link (according to the vertex from which it emanates), and $r$ the representation.
  In our $SU(2)$ case, the fundamental representation is $r=\frac{1}{2}$ and we shall suppress the index $r$ in this case.
 Thus $U_n \equiv U_n^{1/2}$ are $2 \times 2$ matrices of operators.
%

Non-ablian fields generally carry color charges, and hence, unlike in the abelian case, each link carries {\em two} different electric color fields,  the left  and right field characterized by the operators $\left\{L_{n,a}\right\},\left\{R_{n,a}\right\}$.  Their difference along a link can be interpreted as the color charge carried by it. These are, in fact, the left and right generators of the group, and hence they must satisfy the (matrix) algebra of $SU(2)$ \footnote{Or more explicitly: $\left[L_a,\left(U^r\right)_{kl}\right] = \left(\lambda_a^r U^r\right)_{kl}$, etc.},
\begin{equation}
\left[L_a, U^r \right] = \lambda_a^r U^r
\; ; \; \left[R_a, U^r\right] =  U^r \lambda_a^r
\label{algebras1}
\end{equation}
where $\left\{\lambda^r_a\right\}$, the representation matrices, are $\lambda_a^{r=1/2} = \frac{1}{2}\sigma_a$.
As generators of $SU(2)$, the left and right electric fields satisfy the algebra
\begin{equation}
\left[L_{n,a},L_{n,b}\right] = -i \epsilon_{abc}L_{n,c}  \; ; \; \left[R_{n,a},R_{n,b}\right] = i \epsilon_{abc}R_{n,c}
\label{algebras2}
\end{equation}
The left and right generators can be shown to commute with each other, and thus give rise to the same Casimir operator, $\mathbf{L}^2 = \underset{a}{\sum}L_aL_a = \mathbf{R}^2$. Thus, in $SU(2)$ the gauge field Hilbert space on a single link is characterized by three different integer quantum numbers, $j,m,m'$, satisfying
\begin{multline}
\mathbf{L}^2\left| jmm' \right\rangle = \mathbf{R}^2\left| jmm' \right\rangle = j \left(j+1 \right)\left| jmm' \right\rangle \\
L_z \left|jmm'\right\rangle = m \left|jmm'\right\rangle \; ; \;
R_z \left|jmm'\right\rangle = m' \left|jmm'\right\rangle
\label{Hilbert}
\end{multline}

Finally, the $r=1/2$ matter fields are introduced as 2-component spinors $\psi_n$ defined at the vertices. The local color charges are defined by
$Q_{n,a} = \frac{1}{2}\underset{k,l}{\sum}\psi_{n,k}^{\dagger}\left(\sigma_a\right)_{kl}\psi_{n,l}$.
The local gauge invariance is manifested by the conservation of Gauss's law at each vertex, $L_{n,a} - R_{n-1,a} = Q_{n,a}$, for each group index $a$ separately. In local gauge transformations, one picks a group element $V_n$ for each vertex, and acts with it on the gauge and matter fields:
$\psi_n \rightarrow V_n\psi_n$ ; $U_n \rightarrow V^{\dagger}_n U_n V_{n+1}$.
This is a transformation in "group space", i.e. on group indices, and thus, in a gauge invariant Hamiltonian,
all the group space indices (of matter and gauge fields) must be fully contracted (effectively 'traced out'). Thus the simplest "pure gauge" terms are of the form
$\propto \text{Tr}_{\text{group}}\left(U_1U_2U^\dagger_3U^\dagger_4\right)$, where the product is of group elements around a plaquette. Such terms give rise to the
propagating effects in $d+1$ dimensions where $d>1$, but are absent in 1+1-d, where the only possible interactions are gauge-matter ones.
In the following we shall use the staggered fermions method in 1+1 dimensions \cite{Banks1976,Susskind1977,Hamer1977}, where the gauge invariant Hamiltonian
is
\begin{equation}
H = \underset{n}{\sum} \left( \frac{g^2}{2}\mathbf{L}_n^2 + m(-1)^{n} \psi^{\dagger}_n \psi_n + {i\beta}\left(\psi^{\dagger}_n U_n \psi_{n+1} - h.c. \right) \right)
\label{Hamiltonian}
\end{equation}
where $g$ is the theory's coupling constant and $m$ is the fermions mass. We shall denote by $\left|vac\right\rangle$ the zeroth order ground state in the strong coupling limit ($g^2 \gg \beta$, where the $\beta$ part is treated as a perturbation). This state satisfies $\left|vac\right\rangle = \underset{\text{links}}{\bigotimes}\left|000\right\rangle\underset{\text{vertices}\;n}{\bigotimes}\left|\psi_n^{\dagger}\psi_n = 1-(-1)^n\right\rangle$.
The double filling in the odd vertices ("negative mass" vertices) is the "Dirac Sea"; In the continuum limit of staggered fermions, two neighboring two-component spinors become a single four-component one.

In order to simulate this Hamiltonian, one has to find an appropriate realization for the group elements $\left\{U_n\right\}$ and generators $\left\{L_{n,a}\right\},\left\{R_{n,a}\right\}$ fulfilling the unitarity and algebra demands (\ref{algebras1},\ref{algebras2},\ref{Hilbert}).
This can be done using the Jordan-Schwinger map \cite{Jordan,Schwinger}, connecting  harmonic oscillators (bosons) and angular momentum. Mapping between $SU(N)$ and bosonic systems \cite{CoherentSU3,CoherentSUN} allows to express the gauge field operators using bosonic atoms in the prepotential method \cite{MathurSU2,Anishetty}. In this method, for $SU(2)$, one defines four bosonic species on each link $n$: the first two are identified by the operators $a_1,a_2$ on the left side, and the other two $b_1,b_2$ on the right side, constrained by $N_L = N_R$, where $N_L \equiv a_1^{\dagger}a_1 + a_2^{\dagger}a_2$ and $N_R \equiv b_1^{\dagger}b_1 + b_2^{\dagger}b_2$. Then, one can express the unitary operators on each link as $U=U_L U_R$, where
\begin{equation}
U_L = \frac{1}{\sqrt{N_L + 1}}\left(
        \begin{array}{cc}
          a_1^{\dagger} & -a_2 \\
          a_2^{\dagger} & a_1 \\
        \end{array}
      \right)
 \; ; \;
U_R = \left(
        \begin{array}{cc}
          b_1^{\dagger} & b_2^{\dagger} \\
          -b_2 & b_1 \\
        \end{array}
      \right)
      \frac{1}{\sqrt{N_R + 1}}
\end{equation}
Then, by identifying $j = \frac{N_L}{2} = \frac{N_R}{2}$, the generators
\begin{equation}
L_a = \frac{1}{2}\underset{k,l}{\sum}a_k^{\dagger}\left(\sigma_a\right)_{lk}a_l
 \; ; \;
R_a = \frac{1}{2}\underset{k,l}{\sum}b_k^{\dagger}\left(\sigma_a\right)_{kl}b_l
\label{angmom}
\end{equation}
and $\mathbf{L}^2 = \frac{N_L}{2}\left(\frac{N_L}{2} + 1\right)$, $\mathbf{R}^2 = \frac{N_R}{2}\left(\frac{N_R}{2} + 1\right)$,
equations (\ref{algebras1},\ref{algebras2},\ref{Hilbert}) follow.

\begin{figure}
\includegraphics[scale=0.6]{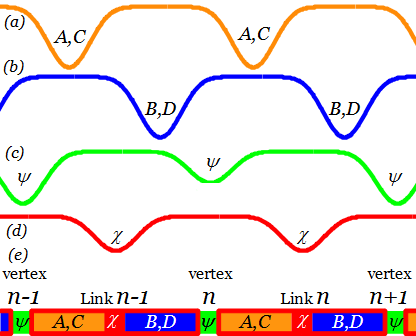}
\caption{A schematic representation of the required bosons (a,b) and fermions (c,d) superlattice structure; (e) shows schematically the combination of all the species to form the links-vertices structure. We have exaggerated the distance between the wells in order to make the structure more clear.}
\label{fig}
\end{figure}

\emph{Pure Gauge Simulation.}
First, we shall discuss the simulation of the gauge field, disregarding the fermions. The simulating system required for that is a set of optical lattices \cite{Lewenstein2012}. Each minimum can contain two different (bosonic) atomic species out of four, $A_{1,2}$ or $B_{1,2}$.  The $A,B$ minima alternate (see Fig. \ref{fig}a,b). We assume that the energy levels of the bosonic modes on each minima are fairly separated, such that we can consider only the lowest one. Since nearest neighbor minima cannot contain similar atomic species, tunneling is eliminated, and thus the only remaining interactions are within the minima - scattering and number terms. Tuning the optical parameters and the chemical potential properly, one gets the Hamiltonian
\begin{equation}
H_E = \frac{g^2}{4} \underset{n}{\sum}\left(\frac{N_{L,n}}{2}\left(\frac{N_{L,n}}{2} + 1\right)+\frac{N_{R,n}}{2}\left(\frac{N_{R,n}}{2} + 1\right)\right)
\end{equation}
but since $\left[H_E,N_{L,n}-N_{R,n}\right] = 0$, if the constraint $N_{L,n}=N_{R,n}$ is initially fulfilled at all links, we get the desired Hamiltonian,
$H_E = \frac{g^2}{2} \underset{n}{\sum}\mathbf{L}_n^2$.
This corresponds to a 1+1 dimensional $SU(2)$ pure-gauge theory. However, since this system has no dynamics at all, we would like to introduce some dynamic color charges (fermions).

\emph{Dynamic Fermions Simulation.}
First, we show how to realize the fermionic mass term. In order to do that, we introduce two-component spinors $\psi_n$ at every vertex (see Fig. \ref{fig}c), i.e. to the left of the $A$ bosonic minima, with an alternating chemical potential, yielding the requested Hamiltonian $H_m = m\underset{n}{\sum}(-1)^{n} \psi^{\dagger}_n \psi_n$. This requires the use of a superlattice, as indicated in Fig. \ref{fig}c.
Next, we introduce the nontrivial interaction term. In order to do that, we introduce ancillary fermionic species, "sitting" in the middle of the links, to the right of the $A$ bosonic minima (see Fig. \ref{fig}d). On each such "virtual vertex" we define a two-component spinor, $\chi_n$, with the local Hamiltonian $H_{\chi} = \lambda\underset{n}{\sum}\chi^{\dagger}_n \chi_n$.
Besides these spinors, we introduce more bosonic species, $C_{1,2}$ and $D_{1,2}$, whose minima overlap with the ones of $A_{1,2}$ and $B_{1,2}$ respectively (see Fig. \ref{fig}a-b), serving as reference baths, all prepared in an identical coherent (i.e. BEC) state $\left|\alpha\right\rangle$, where $\alpha \in \mathbb{R}$, $\alpha \gg 1$.

The required boson-fermion interaction is obtained as a boson-assisted tunneling, where fermionic tunneling is accompanied by an internal boson change, specifically
\begin{multline}
H_f = \frac{\epsilon}{2^{1/4}\alpha} \underset{n,i,j}{\sum} \Bigg(\left(\psi_n^{\dagger}\right)_i\left(\tilde W_{L,n}\right)_{ij} \left(\chi_n\right)_j + \\ + \left(\chi_n^{\dagger}\right)_i\left(\tilde W_{R,n}\right)_{ij} \left(\psi_{n+1}\right)_j + h.c. \Bigg)
\label{ints}
\end{multline}
where
\begin{equation}
\tilde W_L = \left(
        \begin{array}{cc}
          a_1^{\dagger}c_1 & -a_2c^{\dagger}_2 \\
          a_2^{\dagger}c_2 & a_1c^{\dagger}_1 \\
        \end{array}
      \right)
 \; ; \;
\tilde W_R = \left(
        \begin{array}{cc}
          b_1^{\dagger}d_1 & b_2^{\dagger}d_2 \\
          -b_2d^{\dagger}_2 & b_1d^{\dagger}_1 \\
        \end{array}
      \right)
\end{equation}
The overlapping of the Wannier functions  of the different atomic species (see Fig. \ref{fig}) naturally gives rise to such interactions. The minima of the same species are always separated by minima of other ones, and hence tunneling between neighboring same-species minima are unlikely. Thus, only the scattering processes contribute.

In order to eliminate undesired processes and keep only the needed ones, one has to carefully select the hyperfine levels of each of the atomic species such that only the desired processes would conserve angular momentum. Thus, gauge invariance will be a fundamental property of the system (and not an effective one), arising from the natural atomic angular momentum conservation. There are many possible choices for the magnetic levels to be used. If one, for example, chooses $^{7}$Li atoms for the bosons  and $^{40}$K atoms for the fermions, there are many choices of $m_F$ values that should work (the selection rules we consider depend only on the magnetic level $m_F$). A possible choice is
$m(A_1) = 3 ,m(A_2) = 2,m(B_1) = -3 ,m(B_2) = -2 ,m(C_1) = 1,m(C_2) = -3, m(D_1) = -1 ,m(D_2) = 3 ,m(\psi_1)= 3/2 ,m(\psi_2)= -3/2 ,m(\chi_1)= 7/2,m(\chi_2)= -7/2$ \footnote{State dependent potentials for different hyperfine levels require laser detunings relatively close to resonance, which may lead to spontaneous emission over longer time scales. In order to avoid spontaneous emission one may
use other atomic species.}. A detailed discussion can be found in the supplementary part \cite{sup}.

Another term which conserves angular momentum and arises in the proposed setup is $\tilde H_f = \gamma \underset{n}{\sum} \psi_n^{\dagger}\psi_{n} \left(N_{R,n-1} + N_{L,n}\right)$,
but this can be eliminated by a proper choice of the Hamiltonian parameters, as is explained later. Scattering terms of the form $\chi^{\dagger}_n\chi_n\left(N_{L,n}+N_{R,n}\right)$ are also angular momentum conserving and thus possible. However, they vanish in the relevant subspace, $\mathcal{M}$, where the dynamics takes place (see below). Finally, the only angular momentum allowed $\psi-\chi$ and $\chi-\chi$ scattering processes vanish in $\mathcal{M}$. The local $\psi-\psi$ scattering leads to a constant (total number of fermions) in $\mathcal{M}$.


Since $c_i\left|\alpha\right\rangle = \alpha\left|\alpha\right\rangle$, and $c^{\dagger}_i\left|\alpha\right\rangle \approx \alpha\left|\alpha\right\rangle$ (because $\alpha \in \mathbb{R}$, $\alpha \gg 1$, and the same applies for $d_i$), we get that effectively, within our subspace of interest, we can replace $\tilde W_L, \tilde W_R$ with $\alpha W_L, \alpha W_R$, where $W_L = \sqrt{N_L+1}U_L$ and $W_R = U_R\sqrt{N_R+1}$. Thus, one effectively gets
\begin{multline}
H_f = \frac{\epsilon}{2^{1/4}} \underset{n,i,j}{\sum} \Bigg(\sqrt{N_{L,n}+1}\left(\psi_n^{\dagger}\right)_i\left(U_{L,n}\right)_{ij} \left(\chi_n\right)_j + \\ + \left(\chi_n^{\dagger}\right)_i\left(U_{R,n}\right)_{ij} \left(\psi_{n+1}\right)_j\sqrt{N_{R,n}+1} + h.c. \Bigg)
\end{multline}
Note that this Hamiltonian is gauge-invariant, if we take into account the $\chi$s as well when doing a gauge transformation.

Next, assume that $H_{\chi}$ is the largest energy scale, i.e. $\lambda \gg g^2,m,\epsilon$. In that case, we can treat $H_E+H_m+H_f + \tilde H_f$ as perturbations to the large constraint $H_{\chi}$. If we prepare initially the system with no $\chi$ atoms at all, we adiabatically eliminate them and obtain an effective Hamiltonian expansion in this subspace \cite{Soliverez1969}, which we denote by $\mathcal{M}$. In first order, we get $H_E + H_m + \tilde H_f$.
In second order, there are several possible contributions. Each virtual action of $H_f$ requires a second operation of it on the same link, in order to return to $\mathcal{M}$. The first type of processes yield the Hamiltonian
$H_f' = -\frac{\epsilon^2}{\sqrt{2}\lambda\gamma} \tilde H_f$ (see the supplemental part \cite{sup} for details). Choosing $\frac{\epsilon^2}{\sqrt{2}\lambda}=\gamma$, we eliminate both $\tilde H_f, H_f'$ as anticipated above.

The other type of processes are of utmost interest. There, a fermion hops to the middle of a link, and then makes its way to the other side. Mathematically (details in the supplemental part \cite{sup}), one gets
$H_{\beta} = \frac{\beta}{\sqrt{2}} \underset{n}{\sum}\left(\psi_{n}^{\dagger}\sqrt{N_{L,n}+1}U_n\sqrt{N_{R,n}+1}\psi_{n+1} + h.c. \right)$
 where $\beta \equiv -\frac{\epsilon^2}{\lambda}$. This yields the full effective Hamiltonian, to second order, $H_{eff} = H_E + H_m + H_{\beta}$.
The gauge-fermions coupling is folded within $H_{\beta}$. Now, let us perform on all the "real" fermions the canonical transformation
$\psi_n \rightarrow i^n \psi_n$.
This transformation leaves $H_m$ invariant, but introduces the missing complex factors to $H_{\beta}$.

A single action of $H_{\beta}$ on $\left|vac\right\rangle$ raises the flux on the relevant link to $j=\frac{1}{2} (N_L=N_R=1)$. Since a link in its ground state contains no $A,B$ bosons, only the creation operators part of $U$, $U_+$, would contribute, and one would get, before its operation, $\sqrt{N+1} \left|vac\right\rangle = \left|vac\right\rangle$, and after its action, $\sqrt{N+1} U \left|vac\right\rangle = \sqrt{N+1} U_+ \left|vac\right\rangle = \sqrt{2} U_+ \left|vac\right\rangle = \sqrt{2} U \left|vac\right\rangle$. Thus we conclude that
$H_{\beta}\left|vac\right\rangle = i{\beta}\underset{n}{\sum}\left(\psi^{\dagger}_n U_n \psi_{n+1} - h.c. \right)\left|vac\right\rangle$ as desired.

One can also show (as done in the supplemental part \cite{sup}) that a double operation of $H_{\beta}$ on $\left|vac\right\rangle$ is
equivalent to the simulated system's case as well, since the flux never exceeds $j=1/2$ (if initially $j \leq 1/2$ everywhere.
 Furthermore, all the gauge invariant states (with $j=1/2$ charges) can be obtained by operations of the $H_{\beta}$ terms on $\left|vac\right\rangle$. Thus, we conclude that if the system did not contain initially any any links with $j>\frac{1}{2}$, this of course also applies for higher orders, which can be obtained by iterating $H_{\beta}$ several times. Therefore, if the system does not contain initially any links with $j>\frac{1}{2}$, one can effectively drop the square roots of number operators and write
\begin{equation}
H_{\beta} = {i\beta}\underset{n}{\sum}\left(\psi^{\dagger}_n U_n \psi_{n+1} - h.c. \right)
\end{equation}
as in (\ref{Hamiltonian}), and get an accurate simulation for \emph{all orders} in $H_{\beta}$ and all the parameters' regimes.

\emph{Initial state preparation and possible measurements.}
Initially, one can prepare the system in the ground state of the strong limit, $\left|vac\right\rangle$, in which the $C,D$ atoms are in the $\left|\alpha\right\rangle$ state previously defined, and there are no other bosons all around the lattice. The fermions should be prepared as explained before. The fermion interactions should be switched off, i.e. $\beta=0$ or $\epsilon = 0$, for example, by deepening the optical lattice minima. One can then create, using single addressing lasers \cite{Bakr2009,Weitenberg2011}, charges with the appropriate flux tubes connecting them. One could either create mesons, whose length should be odd (since we are using staggered fermions and "quarks" and "anti-quarks" are on alternating vertices) or baryons, sitting on the same vertex, as described in \cite{Hamer1977}. As long as no states with $j>\frac{1}{2}$ on any link are produced, the dynamics would be exactly as in the simulated model. Then the fermionic dynamics can be turned on. Afterwards, one can change the parameters $m$, $g$, and if it is done adiabatically, one can go from the strong to the weak limit and see the consequences (no phase transition is expected in 1 spatial dimension, of course).
Measurements can be done by probing the number of "real" fermions and $A,B$ bosons over the lattice. Another possibility is to realize the Wilson-Loop area law measurement, proposed in \cite{Topological,Dynamic} for a non-abelian system.

Before concluding, we shall emphasize again that in this model gauge invariance is fundamental and exact, inherited from angular momentum conservation. This makes the model robust against errors and corrections: as the symmetry is already manifested in the basic Hamiltonian (\ref{ints}), the system cannot leave the gauge invariant subspace.

\emph{Generalization to more dimensions.} As mentioned before, lattice gauge theories in $d+1$ dimensions $(d>1)$ contain "magnetic" plaquette terms, unlike the $1+1$ dimensional case presented here. Thus the generalization to higher dimensions requires more complicated techniques \cite{future}.
 Another important consequence of these terms is that for $d>1$ the $\sqrt{N+1}$ operators encountered here cannot be avoided, since these closed flux loops do not allow remaining in the $j\leq \frac{1}{2}$ subspace, or, phrased differently,
  since the flux is no longer confined in a line, it can spread and raise $j$. In this work we have considered the simpler 1+1 dimensional model, although some of the ideas introduced here may be used as a basis to build higher dimensional versions.

  Finally, let us emphasize that it is fair to recognize that the simulation proposed here requires setups that are much more complex than the ones required for the simulation of condensed matter models, something which will make the realization of the present proposal extremely challenging. However, the purpose of our work is to show that this, in principle, is possible, and to trigger the development of the required experimental techniques, since
  simulation of gauge theories may have a strong impact well beyond atomic and condensed matter physics.

After the completion of this work, two other proposals for quantum simulations of non-abelian gauge theories with cold atoms have been suggested, for a strong-coupling Rishon-Link model \cite{Rishon2012} and an $SU(2)$ gauge magnet \cite{Tagliacozzo2012}.

\emph{Acknowledgements.}
BR acknowledges the support of the Israel Science Foundation, the German-Israeli Foundation, and the European Commission (PICC).
IC is partially supported by the EU project AQUTE.
EZ acknowledges the support of the Adams Fellowship, of the Israel Academy of Sciences and Humanities.

\bibliography{ref}

\section{Supplemental material}

\subsection{Gauge invariance and angular momentum conservation}
Our model requires the creation of interactions of the form
\begin{multline}
H_f = \frac{\epsilon}{2^{1/4}\alpha} \underset{n,i,j}{\sum} \Bigg(\left(\psi_n^{\dagger}\right)_i\left(\tilde W_{L,n}\right)_{ij} \left(\chi_n\right)_j + \\ + \left(\chi_n^{\dagger}\right)_i\left(\tilde W_{R,n}\right)_{ij} \left(\psi_{n+1}\right)_j + h.c. \Bigg)
\end{multline}
where
\begin{equation}
\tilde W_L = \left(
        \begin{array}{cc}
          a_1^{\dagger}c_1 & -a_2c^{\dagger}_2 \\
          a_2^{\dagger}c_2 & a_1c^{\dagger}_1 \\
        \end{array}
      \right)
 \; ; \;
\tilde W_R = \left(
        \begin{array}{cc}
          b_1^{\dagger}d_1 & b_2^{\dagger}d_2 \\
          -b_2d^{\dagger}_2 & b_1d^{\dagger}_1 \\
        \end{array}
      \right)
\end{equation}
as explained, such interaction term is obtainable if one takes advantage of hyperfine angular momentum conservation.
The key point is that on each bose-fermi interaction, one boson-fermion couple is created instead of another pair, which is annihilated. Angular momentum conservation (which is a kind of rotating-wave approximation)
forces the total $m_F$ of the created pair to equal the $m_F$ of the annihilated one. Thus, one should specifically choose the hyperfine levels ($m_F$s) of the various atomic species such that only the desired processes can occur.

This is done the following way. First, let us look at the interactions on the left side of each link, which involve four bosonic species ($A_{1,2},C_{1,2}$) and four fermionic species ($\psi_{1,2},\chi_{1,2}$). The $m_F$ values should be picked such that: (1) The total $m$ value of a pair created in one interaction will exactly equal the total $m$ of the pair annihilated in this interaction, and not equal the total $m$ in any other boson-fermion pair in the system; (2) The total $m$ value of any pair not participating in the desired interactions does not equal the total $m$ of any other pair, either participating in the interactions or not.

After fulfilling this requirement, one has to examine the right side interactions. The fermionic quantum numbers are now inherited from the left side, but the four other bosonic species ($B_{1,2},D_{1,2}$) have to be seeked in the same manner, following the same two requirements.

The entire process can be described using two summation tables, for the left and right sides, with the fermions and bosons $m$ values as the column and row headings. The required Hamiltonian terms determine which four pairs of elements of each of the tables must be equal, and all the other elements must be unique. Thus, due to angular momentum conservation, one gets only the required terms, and the non-species-changing scattering described by $H_f$ in the paper, which is later cancelled by $H_f'$. As explained in the paper, most of the other terms do not contribute  inside $\mathcal{M}$.

The following tables present the summation tables for the choice of $m$s given in the paper. The colors match the proper elements of the $\tilde W_L, \tilde W_R$ matrices, given below.

\begin{widetext}

\subsubsection{Left processes example}
\begin{equation}
\tilde W_L = \left(
        \begin{array}{cc}
          \textcolor{red}{a_1^{\dagger}c_1} & \textcolor{green}{-a_2c^{\dagger}_2} \\
          \textcolor{blue}{a_2^{\dagger}c_2} & \textcolor{cyan}{a_1c^{\dagger}_1} \\
        \end{array}
      \right)
\end{equation}

\begin{center}
\begin{tabular}{ | c | c | c | c | c |}
  \hline
    & $m(A_1) = 3$ & $m(A_2) = 2$ & $m(C_1) = 1$ & $m(C_2) = -3$ \\  \hline
  $m(\psi_1)= 3/2$ & \textbf{\textcolor{red}{$9/2$}} & $7/2$ & $5/2$ & \textbf{\textcolor{green}{$-3/2$}} \\  \hline
  $m(\psi_2)= -3/2$ & $3/2$ & \textbf{\textcolor{blue}{$1/2$}} & \textbf{\textcolor{cyan}{$-1/2$}} & $-9/2$ \\  \hline
  $m(\chi_1)= 7/2$ & $13/2$ & $11/2$ & \textbf{\textcolor{red}{$9/2$}} & \textbf{\textcolor{blue}{$1/2$}} \\  \hline
  $m(\chi_2)= -7/2$ & \textbf{\textcolor{cyan}{$-1/2$}} & \textbf{\textcolor{green}{$-3/2$}} & $-5/2$ & $-13/2$ \\
  \hline
\end{tabular}
\end{center}

\subsubsection{Right processes example}

\begin{equation}
\tilde W_R = \left(
        \begin{array}{cc}
          \textcolor{red}{b_1^{\dagger}d_1} & \textcolor{green}{b_2^{\dagger}d_2} \\
          \textcolor{blue}{-b_2d^{\dagger}_2} & \textcolor{cyan}{b_1d^{\dagger}_1} \\
        \end{array}
      \right)
\end{equation}

\begin{center}
\begin{tabular}{ | c | c | c | c | c |}
  \hline
    & $m(B_1) = -3$ & $m(B_2) = -2$ & $m(D_1) = -1$ & $m(D_2) = 3$ \\  \hline
  $m(\psi_1)= 3/2$ & $-3/2$ & \textbf{\textcolor{blue}{$-1/2$}} & \textbf{\textcolor{red}{$1/2$}} & $9/2$ \\  \hline
  $m(\psi_2)= -3/2$ & \textbf{\textcolor{cyan}{$-9/2$}} & $-7/2$ & $-5/2$ & \textbf{\textcolor{green}{$3/2$}} \\  \hline
  $m(\chi_1)= 7/2$ & \textbf{\textcolor{red}{$1/2$}} & \textbf{\textcolor{green}{$3/2$}} & $5/2$ & $13/2$ \\  \hline
  $m(\chi_2)= -7/2$ & $-13/2$ & $-11/2$ & \large{\textbf{\textcolor{cyan}{$-9/2$}}} & \large{\textbf{\textcolor{blue}{$-1/2$}}} \\
  \hline
\end{tabular}
\end{center}

\end{widetext}

\subsection{The second order effective Hamiltonian terms}
As explained in the text, in the second order, there are several possible contributions to the effective Hamiltonian. Each virtual operation of $H_f$ requires a second operation of it on the same link, in order to return to $\mathcal{M}$.

In the first type of processes, a "real" fermion $\psi$ hops to the left, over $B$ bosons, to the middle of the link, and then jumps back:
\begin{multline}
-\frac{\epsilon^2}{\sqrt{2}\lambda} \underset{n,i,j,l,m;\left|\phi\right\rangle \notin \mathcal{M} }{\sum}
\Bigg(
\left(\psi_{n+1}^{\dagger}\right)_i \sqrt{N_{R,n}+1} \left(U_{R,n}^{\dagger}\right)_{ij} \left(\chi_n\right)_j \times \\
\left|\phi\right\rangle\left\langle\phi\right|
\left(\chi_n^{\dagger}\right)_l \left(U_{R,n}\right)_{lm} \sqrt{N_{R,n}+1} \left(\psi_{n+1}\right)_m
\Bigg) \\
=-\frac{\epsilon^2}{\sqrt{2}\lambda} \underset{n}{\sum}\left(N_{R,n-1}+1\right)\psi_{n}^{\dagger}\psi_{n}
\end{multline}
and similarly, a "real" fermion $\psi$ can hop to the right, over $A$ bosons, and back, giving rise to a similar contribution. Together they add up to (disregarding a constant part, proportional to the total number of fermions in the system)
\begin{equation}
H_f' = -\frac{\epsilon^2}{\sqrt{2}\lambda} \underset{n}{\sum}\psi_n^{\dagger}\psi_{n} \left(N_{R,n-1} + N_{L,n}\right)
\end{equation}
Note that if by setting $\gamma = \frac{\epsilon^2}{\sqrt{2}\lambda}$ in $\tilde H_f$ , one gets $H_f' + \tilde H_f = 0$ and both this unwanted Hamiltonians cancel each other and may be disregarded.

The second type of processes is of more interest. There, a fermion hops to the middle of a link, and then makes its way to the other side. Mathematically, one gets
\begin{multline}
-\frac{\epsilon^2}{\sqrt{2}\lambda} \underset{n,i,j,l,m;\left|\phi\right\rangle \notin \mathcal{M} }{\sum}
\Bigg(
\left(\psi_{n}^{\dagger}\right)_i \sqrt{N_{L,n}+1} \left(U_{L,n}\right)_{ij} \left(\chi_n\right)_j \times \\
\left|\phi\right\rangle\left\langle\phi\right|
\left(\chi_n^{\dagger}\right)_l \left(U_{R,n}\right)_{lm} \sqrt{N_{R,n}+1} \left(\psi_{n+1}\right)_m
\Bigg) + h.c. \\
=-\frac{\epsilon^2}{\sqrt{2}\lambda} \underset{n}{\sum}\left(\psi_{n}^{\dagger}\sqrt{N_{L,n}+1}U_n\sqrt{N_{R,n}+1}\psi_{n+1} + h.c. \right)
\end{multline}
setting $\beta \equiv -\frac{\epsilon^2}{\lambda}$ one gets $H_{\beta}$ defined in the paper.

\subsection{Second order $H_{\beta}$ actions}
 Here we shall elaborate on the second order action of $H_{\beta}$ on $\left|vac\right\rangle$. The cases involving products of two different link operators are easily deduced to be equivalent to the real simulated theory, as in the single operation case. The ones which involve product of operators of the same link require a more careful consideration. First, consider the terms which involve $U$ and $U^{\dagger}$ on the same link:
\begin{multline}
\frac{\beta ^2}{2} \left(\psi_{n+1}^{\dagger}\right)_i\sqrt{N_{R,n}+1}\left(U_n^{\dagger}\right)_{ij}\sqrt{N_{L,n}+1}\left(\psi_{n}\right)_j \times \\
\left(\psi_{n}^{\dagger}\right)_k\sqrt{N_{L,n}+1}\left(U_n\right)_{kl}\sqrt{N_{R,n}+1}\left(\psi_{n+1}\right)_l
\left|vac\right\rangle
\end{multline}
but note that, as before, the rightmost $\sqrt{N+1}$ is just $1$, and the $N+1$ in the middle is $2$. Moreover,
$\left(\psi_{n}\right)_j \left(\psi_{n}^{\dagger}\right)_k = \delta_{jk} - \left(\psi_{n}^{\dagger}\right)_k \left(\psi_{n}\right)_j$. If $n$ is odd,
$\left(\psi_{n}^{\dagger}\right)_k \left(\psi_{n}\right)_j \left|vac\right\rangle = \delta_{jk} \left|vac\right\rangle$, and we get zero, as in the simulated system. If $n$ is even, $\left(\psi_{n}^{\dagger}\right)_k \left(\psi_{n}\right)_j \left|vac\right\rangle = 0$, and due to the unitarity of $U_n$ we get
\begin{equation}
\beta ^2 \sqrt{N_{R,n}+1} \psi_{n+1}^{\dagger}\psi_{n+1}
\left|vac\right\rangle =
\beta ^2 \psi_{n+1}^{\dagger}\psi_{n+1}
\left|vac\right\rangle
\end{equation}
as in the simulated system as well.

Finally, consider terms of the form
\begin{multline}
\frac{\beta ^2}{2} \left(\psi_{n}^{\dagger}\right)_i\sqrt{N_{L,n}+1}\left(U_n\right)_{ij}\sqrt{N_{R,n}+1}\left(\psi_{n+1}\right)_j \times \\
\left(\psi_{n}^{\dagger}\right)_k\sqrt{N_{L,n}+1}\left(U_n\right)_{kl}\sqrt{N_{R,n}+1}\left(\psi_{n+1}\right)_l
\left|vac\right\rangle
\end{multline}
The three right square roots give us $2$ as well, cancelling the denominator. The first (right) $U$ operates on $\left|vac\right\rangle$, and thus it is $U_+$. The second (left) one contributes $U_+ + U_-$. The $U_-$ contribution lowers the flux on the link to $j=0$, and thus the leftmost $\sqrt{N+1}$ is $1$ and we get an exact equivalence to the simulated system. The $U_+$ contribution vanishes here as well as in the simulated system, due to canonical fermionic commutation relations. Thus, in conclusion, the second order of $H_{\beta}$ is equivalent to the simulated system's case as well.
Note, however, that all the gauge invariant states (with $j=1/2$ charges) can be obtained by operations of the $H_{\beta}$ terms on $\left|vac\right\rangle$. Thus, we conclude that up to second order, $H_{\beta}$ is equivalent to the simulated system, since it never produces states with $j>\frac{1}{2}$, given that the system did not contain initially any any links with $j>\frac{1}{2}$. This of course also applies for higher orders, which can be obtained by iterating $H_{\beta}$ several times. Therefore, if the system does not contain initially any links with $j>\frac{1}{2}$, one can effectively drop the square roots of number operators and write
\begin{equation}
H_{\beta} = {i\beta}\underset{n}{\sum}\left(\psi^{\dagger}_n U_n \psi_{n+1} - h.c. \right)
\end{equation}
as in (4) in the letter, and get an accurate simulation for \emph{all orders} in $H_{\beta}$ and all the parameters' regimes.

\end{document}